\documentclass[10pt]{article}
\usepackage{graphics}
\usepackage[square,numbers,comma,sort&compress]{natbib}
\usepackage[pdftex]{graphicx}
\usepackage{color}
 \usepackage{fullpage}

\usepackage{amssymb,amsfonts,amsmath}
\usepackage[nomarkers]{endfloat}

\title{Dielectric mixtures\\ Importance and theoretical approaches}
\author{Enis Tuncer\\ Dielectrics \& Electrophysics Laboratory\\ General Electric Global Research Center\\ Niskayuna NY 12309 USA}
\begin{document}
\maketitle
\begin{abstract}
Physics of dielectric mixtures are presented to stimulate discussion and to provide information on the recent advances in this topic.


\end{abstract}
\tableofcontents
\section{Introduction}
Many engineering materials used in electrical insulation are composites \citep{DakinChapter}. As \citeauthor{DakinChapter} stated `Composite materials are at least an order of magnitude more diverse and perhaps as much more complex in their behavior than simple one-component materials.' Books and review papers on dielectric mixtures or electrical properties of composites \cite{MiltonBook,SihvolaBook,SahimiBookI,SahimiBookII,TorquatoBook,Bergman1992,BBReview,SteemanTurnhout,Boyd1996,Tuncer2002a,TuncerPhilMagLett,niklasson:3382,Clerc1990} discuss the properties of these systems with little or no focus on in their electrical insulation properties. The importance of the insulation properties would arise  from the frequency dependent dielectric losses and changes in the dielectric breakdown strength of the composite system. The latter case is mainly dependent on the defect density and could be separated in to different branches as stated by \citet{NelsonChapter}; by the motion of free charges, charge injection from the electrodes, charge multiplication and space charge formation, and the dissipation energy in the material (electro-thermal effect). Some of these effects would be temperature activated that can make the analysis cumbersome \cite{Dakin2006,Dakin1956,Dakin1978}. However, they are effected by the local electric field and the energy landscape, which again brings the dielectric properties of the composite. We will here focus on the dielectric relaxation in dielectric mixtures by taking in to account the intrinsic properties of components and the geometrical arrangement of phases \cite{TuncerPhD,TuncerPhilMagLett,Tuncer2002a,Tuncer2001a,TuncerINTECH2011,Tuncer2002b,ma3010585,TuncerJPD2005,TuncerSpectralPRB,Tuncer2005JPCMLET,EyreMilton,mil81,Brosseau1,Brosseau2,Sareni1,sar97,sar97mag}.

\section{Background} 
\begin{figure}[htp]
  \centering
  \includegraphics[width=.8\linewidth]{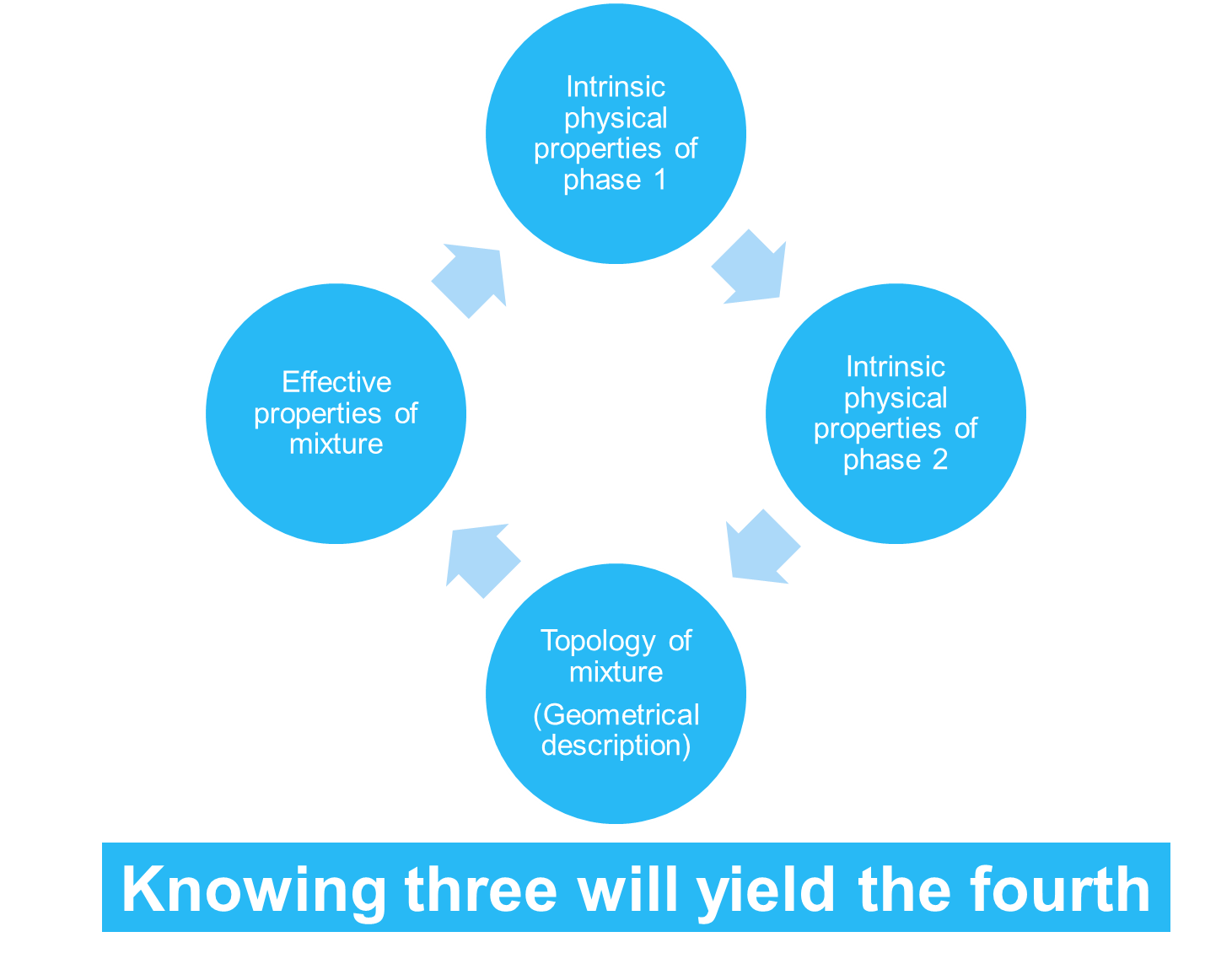}
  \caption{Relationship between different physical parameters in a dielectric mixtures. As mentioned on the bottom of the chart have the information in three of the circles, one can estimate the missing information in the remaining circle.}
  \label{fig:chart}
\end{figure}

The question to answer in designing an insulation system or material is how we can predict the effective electrical properties of the system; notice that similar approaches can be applied to mechanical and thermal properties. We would also like to decompose  contributions of the constituent properties for determination of aging and degradation from non-destructive measurement methods and diagnostic tests. A comprehensive understanding of the electrical properties of insulation systems would be tremendously beneficial for new equipment design and assessing already installed equipment maintenance. A chart regarding the relation ship between material properties is shown in Fig.~\ref{fig:chart}. The presented method later in text would make it possible to obtain the required information in a dielectric mixture, such that one can estimate (i) the effective dielectric permiottivity of a mixture, or (ii) estimate the permittivity of one of the constituents, or (iii) estimate topology of the mixture.

There have been analytical approaches for dielectric mixtures starting from the early days of electro-technical sciences, see \citet{SihvolaBook} for review and numerous dielectric mixture formulas. For example \citet{Maxwellbook} derived the mixture expression for layered binary structure; this was something interesting since the mixture illustrated a new loss peak (or a dielectric dispersion) due to differences in the electrical properties of the constituents, their permittivity and conductivity \cite{Scaife}. \citet{Wagner1913} expended the mixture electrical properties for a system with spherical inclusions. Both \citet{Fricke1924} and  \citet{Sillars1937} improved Wagner's approach to  ellipsoidal shape inclusions. This relaxation due to the mixture was later called as the Maxwell-Wagner-Sillars (MWS) relaxation corresponding to the interface between the constituents. For dilute mixtures the MWS relaxation is narrow, while for complex mixture topologies and high concentration of inclusions the inter-facial relaxation can be broad \cite{TuncerJPD2005}. While analytical models are of great use, they have limitations due to the assumption in their derivation; they can not be applied to many real systems. 

\begin{figure}[htp]
  \centering
  \includegraphics[width=.8\linewidth]{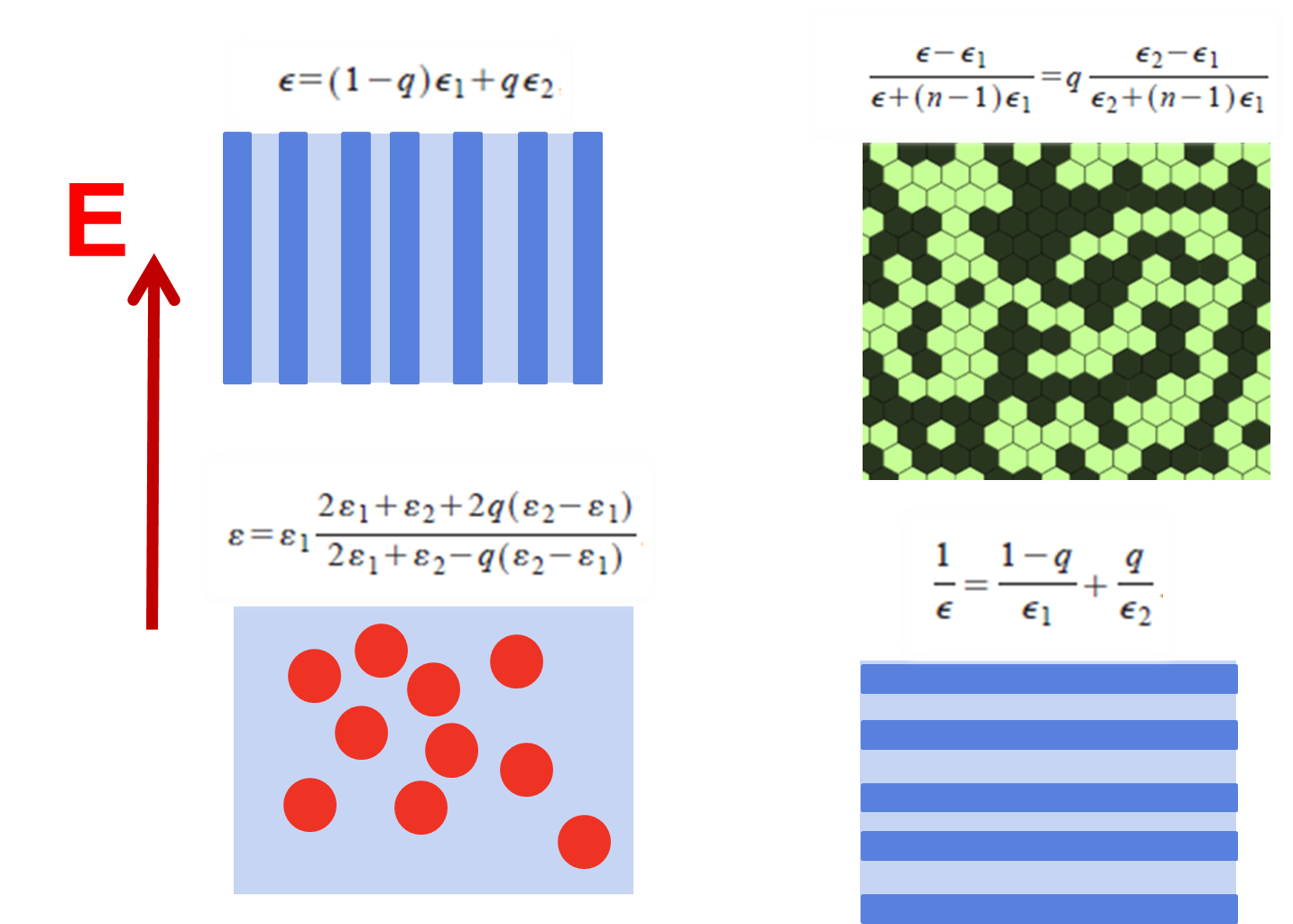}
  \caption{Examples of mixture topologies and corresponding effective dielectric permittivity formulas, see \citet{Tuncer2001a} for details on the expressions.}
  \label{fig:boyd}
\end{figure}

A novel approach to dielectric mixtures was first proposed by \citet{Fuchs1975} and then by \citet{Bergman1978}, which considers a summation rule related to the depolarization factors; it is called the spectral density representation. Several improvements in the method were proposed throughout the years \cite{Bergman1979,Bergman3,bergman1982,Bergman1,Bergman1992,Miltona,Milton1981b,Milton1981,Golden1,Golden2,ma3010585,TuncerPhilMagLett,Tuncer2005JPCMLET}. This approach does not rely on one particular shape of inclusions; it considers distribution of shapes or total topology of the mixture. Numerical approaches to resolve the topology of mixtures using spectral density approach was employed previously \cite{TuncerJPD2005,TuncerSpectralPRB,Cherkaev2003,DayPRL2000,DayPhysB1,DayPhysB2,Day,Stroud1999,Fuchs_a,GhoshFuchs,Pech_PhysRevB.74.035120}. Actually the spectral representation is analogous to the distribution of relaxation times approach in the theory of dielectrics\cite{Tuncer2005JPCMLET,TuncerPhilMagLett}. Examples of mixture formulas were shown by \citet{Boyd1996}, which is a nice illustration of the topologies as shown in Fig.~\ref{fig:boyd}. The upper left and bottom right topologies are for the Wiener bounds \cite{wiener}. The topology on the upper right is for Bruggeman's derivation \cite{Bruggeman1935} which represents a bi-percolating structure. The bottom left topology indicates a mixture with spherical inclusions. The importance of the length scales should not be disregarded in the effective material property calculations, such that the diameter of the inclusions should be much smaller than the wavelength of the applied electric field. For example one can apply the effective medium approximations to nanodielectrics in optical frequecies, however, one cannot apply the same to composites with sub-micron size fillers due to similarities in the length scale of the applied electromagnetic wavelength and particle size dimensions. A list of mixture expressions for spherical inclusions were given by \citet{Widjajakusuma2003319}, which illustrated the variety of formulas for a similar problem.

Numerical simulations of dielectric mixtures have been an approach to study the effective electrical properties of these systems \cite{SihvolaBook,BBReview,Tuncer2002a}. The computational methods of the finite elements and finite differences have been employed numerous times in the literature \cite{Brosseau1,Boudida2,Brosseau2,Boudida,Sareni1,sar97,sar97mag,BrosseauPhysRevE.74.031405,BrosseauPhysRevE71,Serdyuk1306717,Kliem5595551,Beroual971447,blanchard:064101,murugaraj:054304,Karkkainen,SihvolaPol,SihvolaEff,Pekonen1999,Sihvola-ellips,TuncerAcc1,KarkkainenPhD,Tuncer2001a,Tuncer2002b,Tuncer2002elec,TuncerJPD2005,Clauzon,Merrill,CLiu,Sareni1,liu97,Vainas,TuncerFaycalRoss2009,Calame2001JAP,Calame2003,Boggs4591430}. The advantage of numerical methods is that there are no limitation in model parameters except the limitations in computation time; depending on the size of the problem computation can be costly. 

\section{Dielectric mixtures}

Factors effecting the effective properties of a mixture can be listed as follows in a first approximation.
\begin{itemize}
\item Intrinsic electrical properties of the constituents, {\em i.e.}, relative permittivity $\varepsilon_r$, ohmic conductivity $\sigma_0$ and dielectric or conductive relaxation processes represented with complex susceptibility $\chi^*$.
\item Concentration of phase.
\item Distribution of phases; topology of the mixture.
\item Chemical inter-phase between constituents.
\end{itemize}
\citet{Fornes20034993} summarized the issues related to some of the discrepancies between theory and experiments from the mechanical aspect of mixtures, however, they are valid for electrical properties as well. As mentioned the theories consider uniform shape and constant dimensions. While in reality we have poly-dispersed particles with distribution of sizes/lengths/thicknesses. Inclusions are often not unidirectional. Interface between the phases are considered perfect in mixture models, however in reality the interface is not perfect and chemical techniques are employed to improve the interface, such as with employment of surfactants; none is included in the model calculations. The intrinsic properties of the materials are assumed to be isotropic and linear, however in reality these assumptions can not be true. In polymer matrix composites the properties of the polymer are assumed to be isotropic, however, depending on the length scale there will be local changes in the density, crystallinity and chain orientation. Most of the theories do not consider interaction between inclusion particles and particle agglomeration, it is well-known that special care must be given to produce well-dispersed particles in fabricated composites.  

\section{Generalization of effective medium approaches}

\begin{figure}[htp]
  \centering
  \includegraphics[width=.8\linewidth]{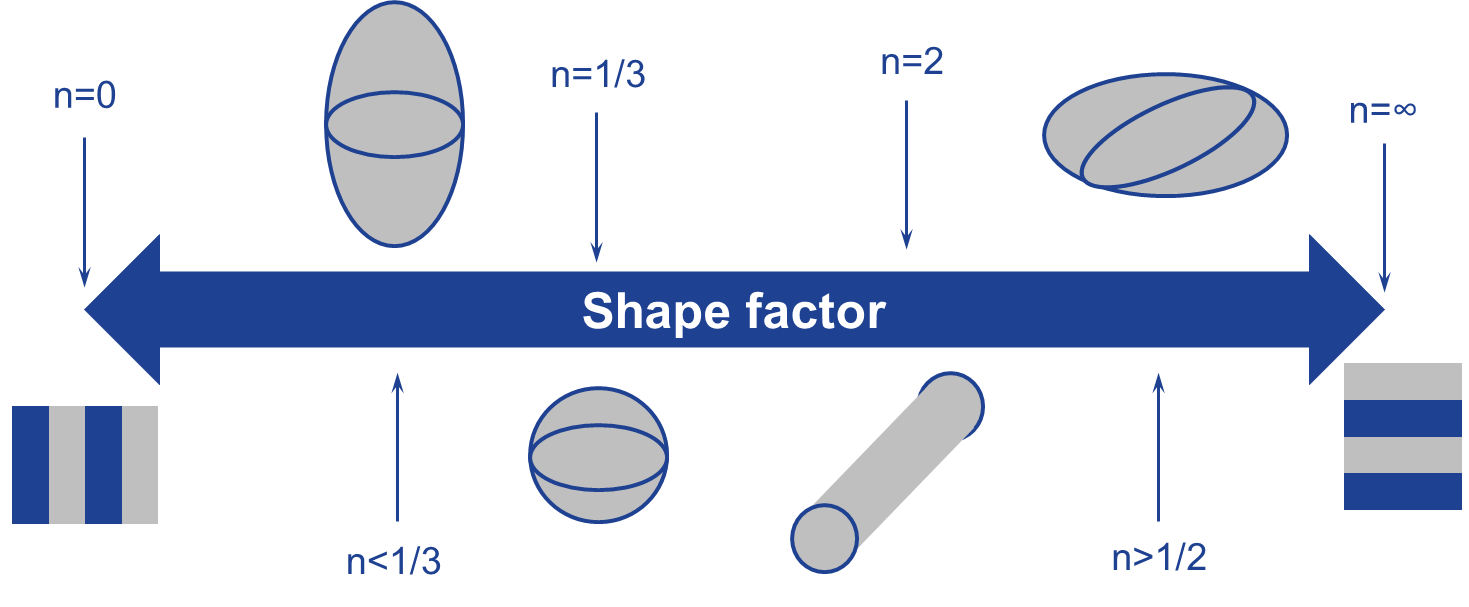}
  \caption{Depolarization factor $\frak{n}$ and shape factor $n$ relationship. The arrow indicates the direction of the electric field. The figure is taken from \citet{TuncerJPD2005}.}
  \label{fig:shape}
\end{figure}

\begin{figure}[htp]
  \centering
  \includegraphics[width=.5\linewidth]{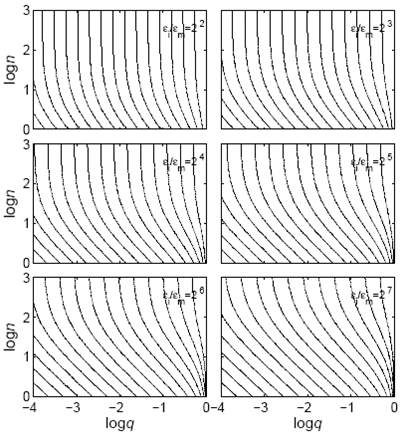}
  \caption{Equi-permittivity plot for a mixture with different relative permittivity ratio of constituents. The shape factor $n$ and concentration $q$ are changed in logarithmic scale. Observe that for large shape factors the relative effective permittivity becomes constant no concentration dependence; straight vertical lines in the plots. While small shape factors together with concentration could yield similar relative effective permittivities.}
  \label{fig:equipermi}
\end{figure}

\begin{figure}[htp]
  \centering
  \includegraphics[width=.5\linewidth]{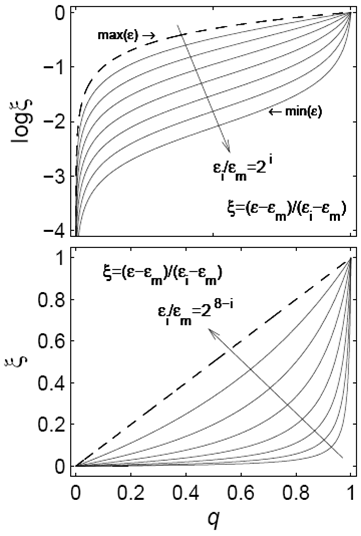}
  \caption{Scaled permittivity $\xi$ as a function of concentration for various ratios of constituent relative permittivities. Phase 2 is considered as spherical particles with $n=3$. The subscripts `m' and `i' represent the matrix and the spherical inclusions. The graph on the top illustrates the the scaled permittivity in logarithmic scale.}
  \label{fig:bounds1}
\end{figure}

Some generalization efforts for the effective medium approaches previously proposed \cite{Pier6,Tuncer2001a}. In these approaches the effective dielectric permittivity of the mixture $\varepsilon_e$ is expressed as
\begin{eqnarray}
  \label{eq:genmixture}
  \frac{\varepsilon_e-\varepsilon_1}{\varepsilon_e+(n-1)\varepsilon_1}=q  \frac{\varepsilon_2-\varepsilon_1}{\varepsilon_2+(n-1)\varepsilon_1}
\end{eqnarray}
where, subscripts `$e$', `$1$' and `$2$' denote the effective, phase 1 and phase 2; the concentration of phase 2 is represented with $q$; $n$ is the shape factor. This expression reduces to the Wiener bounds with $n=1$ and $n=\infty$ for parallel laminates and perpendicular laminates to electric field. The relationship between various shapes  and $n$ are shown in Fig.~\ref{fig:shape}. Spheres and cylinders yield $n=3$ and $n=2$, respectively. The one thing that is important with this expression is that one can obtain the similar relative effective permittivity $\varepsilon_e$ with different combination of $n$ and $q$, which is represented in Fig.~\ref{fig:equipermi}. It is clear that for design purposes one have the option to select specific filler (permittivity of phase), concentration and shape of inclusion phase to optimize for the required permittivity value. For example for large discrepancy in permittivities the region of interest has a larger area, see bottom right graph in Fig.~\ref{fig:equipermi}. If the relative effective permittivity is represented as the scaled permittivity $\xi$ \cite{TuncerPhilMagLett,Tuncer2005JPCMLET,ma3010585},
\begin{eqnarray}
  \label{eq:scaled}
  \xi=\frac{\varepsilon_e-\varepsilon_1}{\varepsilon_2-\varepsilon_1},
\end{eqnarray}
then the relationship between the ratio of permittivities and the concentration for a constant shape factor can be visualized as in Fig.~\ref{fig:bounds1}. The boundaries in Fig.~\ref{fig:bounds1} indicate the highest and the lowest scaled permittivities that can be achieved with a given set of materials. For large permittivity discrepancies the shape of the curve is like a rotated `L'. Observe that we have not considered particle interaction up to this point, therefore the representation at high filler concentrations would not be accurate.

When frequency dependent properties of constituents are considered the problem is a little bit complicated due to the inter-facial polarization and dispersion coupling between phases, see \citet{TuncerSpectralAPAMSP} for details. It has been shown that the representation of the dielectric data in effective complex resistivity level ($\rho_e^*=(\imath\omega\varepsilon_0\varepsilon_e)^{-1}$), the bounds can be observed clearly, unlike in the effective complex permittivity.

\section{Spectral density representation} 

To go a little bit further with the modeling of mixtures, the spectral density representation will be discussed here \cite{FuchsLiu,Fuchs1975,Bergman1978,Bergman4,Milton1981,Golden2,Golden1,TuncerSpectralAPAMSP,TuncerJPD2005,TuncerSpectralPRB,ma3010585}. Let us assume we have several different shapes of inclusions in the system with known shape parameters $n$, then by using a summation one can add the contributions of these entities to the overall effective permittivity. In a distribution of shapes similar approach would be derived with a distribution of shape factors. Using the scaled permittivity representation, the scaled permittivity $\xi$ is then expressed \cite{TuncerPhilMagLett,Tuncer2005JPCMLET,ma3010585} as
\begin{eqnarray}
  \label{eq:SDR}
  \xi=\xi_p+\int_{0+}^1{\sf g}(x)(1+\varpi x)^{-1} dx.
\end{eqnarray}
Here $x$ and ${\sf g}(x)$ are the spectral parameter and the distribution of spectral parameters; $\varpi$ is the spectral frequency; $\xi_p$ is the concentration of phase 2 that is not contributing to polarization, in other words the percolating strength or the infinite clusters \cite{Percolation}. The function $g(x)$ contains the information related to shape and their concentration so,
\begin{eqnarray}
  \label{eq:gx}
  \xi_n=\int {\sf g}(x) dx.
\end{eqnarray}
The total concentration of phase 2 for example will be the sum of $\xi_p$ and $\xi_n$, which is the non-percolating portion of the materials; the finite clusters\cite{Percolation}. The most expected spectral parameter $\bar{x}$ (shape) can be estimated from ${\sf g}(x)$ as well,
\begin{eqnarray}
  \label{eq:mostgx}
  \bar{x}=\int x {\sf g}(x) dx.
\end{eqnarray}
Examples of application of the method and resolving different distributions were shown in the literature\cite{ma3010585,TuncerJPD2005,TuncerSpectralPRB,Stroud1999,Day,DayPhysB2,DayPhysB1,DayPRL2000,Cherkaev2003,Fuchs_a,GhoshFuchs,Ghosh}. In some cases the distribution of spectral parameters were resolved with individual peaks indicating a discrete structure \cite{Tuncer2008OptComm}. Application of the method as a micro-structural analysis was proposed which might be a good possibility to electrically characterize composites \cite{TuncerDrummy}. However, one needs dielectric property information of the system in advance to perform the analysis, as shown in Fig.~\ref{fig:chart}.
 
\section{Resistor-capacitor network and its spectral density}
\begin{figure}[htp]
  \centering
  \includegraphics[width=.5\linewidth]{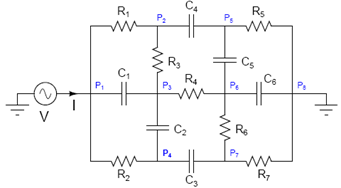}
  \caption{Model circuit to illustrate the construction of the RC network.}
  \label{fig:circuit}
\end{figure}

\begin{figure}[htp]
  \centering
  \includegraphics[width=.5\linewidth]{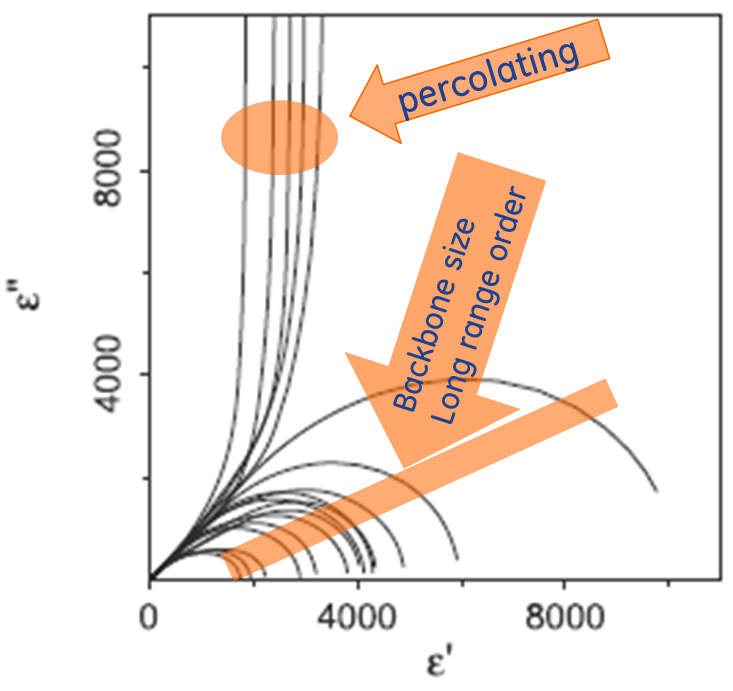}
  \caption{The Argand plot for the complex permittivity estimations for the $512\times512$ resistor-capacitor network with 50\% resistor coverage. The percolating and non-percolating structures could be seen clearly with those percolating having large dielectric losses. The non-percolating structure indicate a clear resistive or dielectric response with long range order.}
  \label{fig:colecole}
\end{figure}

To better illustrate the strength of the spectral density analysis results from an resistor-capacitor circuit modeling were studied. For system and modeling information, please refer to  \citet{TuncerFaycalRoss2009}. In the analysis different sizes of circuits composed of 50-50 resistor and capacitor occupancy were simulated. The sizes of the circuits were $2^j\times2^j$ with $j=\{4,5,6,7,8,9\}$. Example of a $3\times3$ circuit is shown in Fig.~\ref{fig:circuit}. The dielectric relaxation of the system with  randomly arranged resistors and capacitors show very different responses depending on the backbone and percolation paths. The responses from several circuit cases studied with $512\times512$ lattice size are shown in Fig.~\ref{fig:colecole}. Depending on the structure of the systems, some indicate percolation due to the lining up resistors and the others had broad relaxations; none of the cases were similar. 

\begin{figure}[htp]
  \centering
  \includegraphics[width=.5\linewidth]{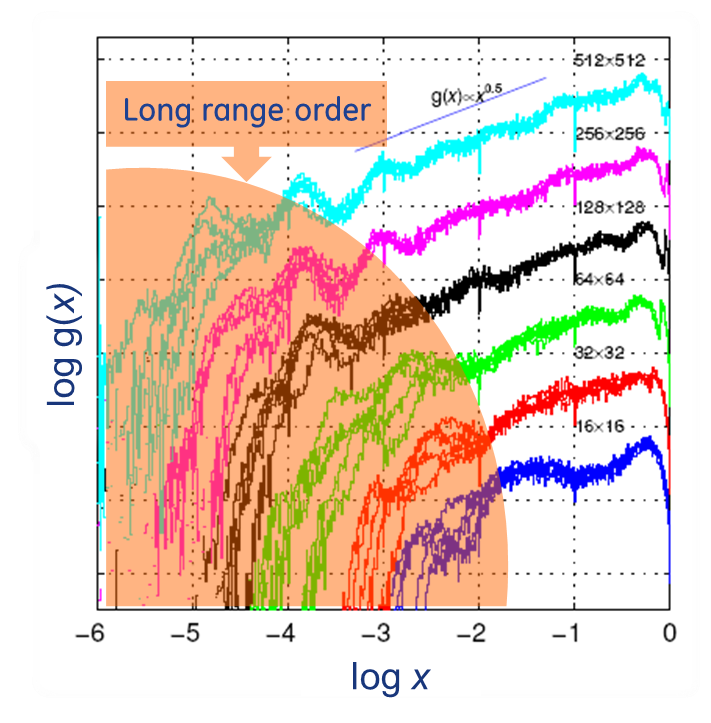}
  \caption{Spectral densities for resistor-capacitor networks with different lattice sizes. Long and short range order information from the analysis are marked in the image. The behavior of the spectral densities are similar for large spectral parameter $x$ values.}
  \label{fig:circspec}
\end{figure}

When we apply the spectral density representation we can see that the randomness and the structure of the spectral densities with respect to considered lattice sizes, Fig.~\ref{fig:circspec}. First the largest lattice considered had the broadest distribution of spectral parameters. This is expected due to the size of the system and the possibilities of generating intricate shapes. While the smallest lattice structure had a narrow distribution. The other striking observation is for large spectral parameters $x$ the structure spectral densities of each set of lattice coincide for the studied cases. While for small spectral parameters  we see differences. This is mainly due to the long range order and increasing size of the lattice generates slight differences in the long range correlations. The short range order is not influenced meaning that the arrangement of the resistors and capacitors would not change the structural information when we consider small lengths scales. One way to apply this observation to dielectric breakdown is in order to fail a material charge species need to correlated with long range order to initiate the breakdown. One interesting example is the electrical treeing where no breakdown is observed until the tree bridges between two electrodes. Taking this example further for those cases where the long range correlation between resistors exists there is a risk to fail the material. Therefore one needs to study the spectral density of the system to determine the existence of any long range correlations. Perhaps it sounds impossible for todays material characterization techniques, but once a library of material properties are build there would be possibilities to study the structure-property relationships in engineering composites.

\section{Bounds on dielectric permittivity: circles of interest}
\begin{figure}[htp]
  \centering
  \includegraphics[width=.5\linewidth]{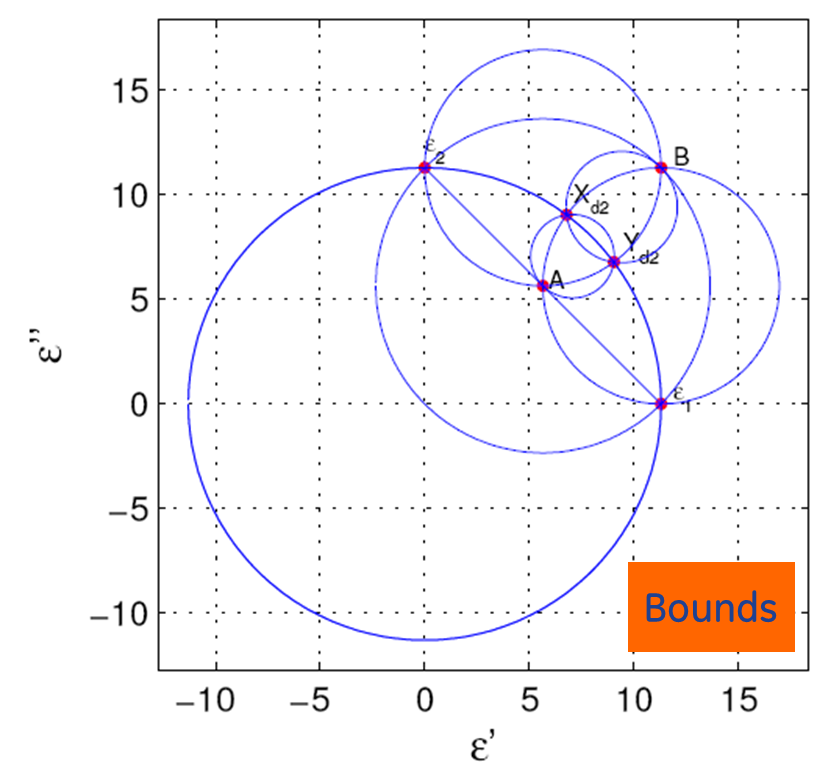}
  \caption{Graphical representation of the bounds for the effective dielectric permittivity of a composite composed of complex permittivities $\varepsilon_1$ and $\varepsilon_2$. The regions of interests are drawn for different values.}
  \label{fig:bounds2}
\end{figure}

\begin{figure}[htp]
  \centering
  \includegraphics[width=1\linewidth]{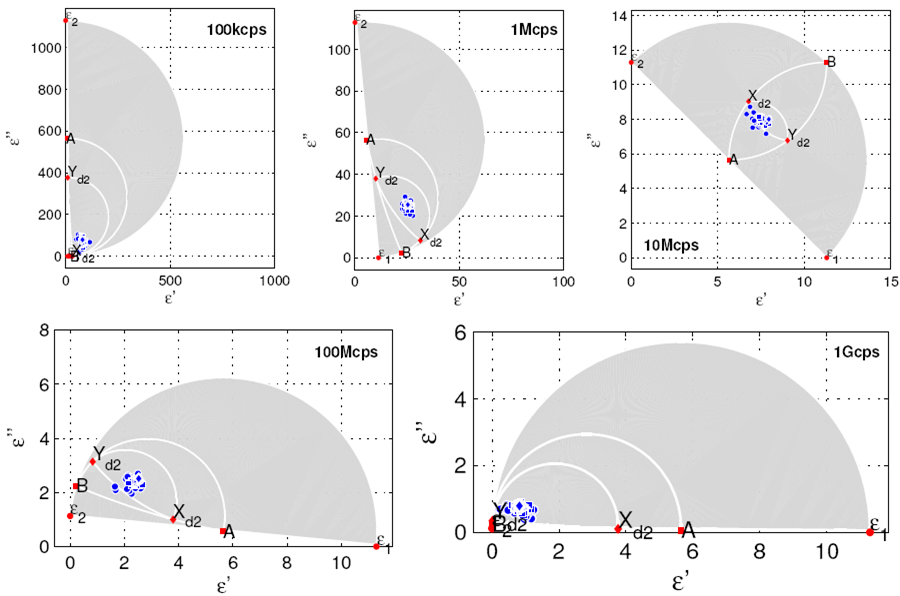}
  \caption{Applications of the bounds to the data of resistor-capacitor networks at different frequencies.}
  \label{fig:circles}
\end{figure}

Before we finish our topic on the dielectric mixtures, I would like to point our attention towards an important analysis technique for predicting or guessing dielectric properties of composites. \citet{milton:300,MiltonBook} has illustrated a graphical technique to study dielectric mixtures. The technique is based on the complex dielectric properties of mixture components. Once an observer gathers information on the composite system it becomes easy to predict the effective properties of the composite. The method was build as follows. It starts with the largest bounds presented by \citet{wiener}, Eq.~(\ref{eq:genmixture}), and includes another one presented by \citet{hashin},
\begin{eqnarray}
  \label{eq:hashin}
  A&\equiv&(1-q)\varepsilon_1+q\varepsilon_2\\
  B&\equiv&((1-q)/\varepsilon_1+q/\varepsilon_2)^{-1}\\
  X_{d_2}&\equiv&\varepsilon_1 + \frac{dq\varepsilon_1(\varepsilon_2-\varepsilon_1)}{d\varepsilon_1+(1-q)(\varepsilon_2-\varepsilon_1)}\\
  Y_{d_2}&\equiv&\varepsilon_2 + \frac{d(1-q)\varepsilon_2(\varepsilon_1-\varepsilon_2)}{d\varepsilon_2+q(\varepsilon_1-\varepsilon_2)}
\end{eqnarray}
Here $d$ is the dimensionality and $q$ is the concentration of phase 2. Observe that the mixture expression of \citet{hashin} for $C$ and $D$ are the same expressions with interchanged phase 1 and phase 2, and $q$ and $(1-q)$. Once can similarly use Eq.~(\ref{eq:genmixture}) for this purpose. Once we have the points $A$ to $Y_{d_2}$ we can now start the graphical analysis as shown in Fig.~\ref{fig:bounds2}. So the electrical properties of the constituents $\varepsilon_1$ and $\varepsilon_2$ determine the absolute limits; $\varepsilon_1A\varepsilon_2$ is the lower bound and $\varepsilon_1B\varepsilon_2$ is the upper bound. Concentration information would yield the positions of $A$ and $B$, then we can narrow the region of interest for the effective permittivity. If we have information on the shape of the inclusions, let us say spherical, then $d=3$, we would get an even narrower region than before marked between $X_{d_2}Y_{d_2}$. Now we can test our regions for the resistor-capacitor network as shown in Fig.~\ref{fig:circles} with dots. For each frequency of interest the limits for the effective permittivity is different due to the complex permittivities or we should the impedance. Since we did not have any information on the size and shape of resistors, some of the points are not inside the smallest region $X_{d_2}Y_{d_2}$. However all the data are located within the region $AX_{d_2}BY_{d_2}$, which indicates that the resistor-capacitor network simulation results were acceptable. 

\section{Summary}

Different approaches to dielectric mixtures are presented with numerous research work related to the topic. The importance of the dielectric properties to the overall insulation design was previously discussed by \citet{DakinChapter}. Here recent approaches and the methods to study dielectric mixtures were shown. No insulation systems were considered because of the number of such systems available. As known by the dielectric mixture community and mentioned by \citet{DakinChapter}, dielectric behavior in composite materials is a complex phenomenon due to involvement of different physical parameters other then the intrinsic properties of the constituents.  However, with novel numerical methods and improved fabrication techniques many of the parameters influence the over all mixture properties could be studied to better control the electrical properties of these systems.  

\section*{Acknowledgment}
The presented work was previously used in the talk on Nanodielectrics Modeling at the Conference on Electrical Insulation and Dielectric Phenomenon held on October 14, 2012 in Montreal Canada. Data from one of an earlier paper were used for the bounds and spectral density estimations. Those data were generated by R. Faycal Hamou at Max-Planck-Institut f{\"u}r Eisenforschung GmbH in 2007. The study and presented data analysis were performed when the author was with Oak Ridge National Laboratory.

\bibliography{f:/labtopbackup/TEX/BibStyles/newref}

% Generated by IEEEtranN.bst, version: 1.13 (2008/09/30)
\begin{thebibliography}{98}
\providecommand{\natexlab}[1]{#1}
\providecommand{\url}[1]{#1}
\csname url@samestyle\endcsname
\providecommand{\newblock}{\relax}
\providecommand{\bibinfo}[2]{#2}
\providecommand{\BIBentrySTDinterwordspacing}{\spaceskip=0pt\relax}
\providecommand{\BIBentryALTinterwordstretchfactor}{4}
\providecommand{\BIBentryALTinterwordspacing}{\spaceskip=\fontdimen2\font plus
\BIBentryALTinterwordstretchfactor\fontdimen3\font minus
  \fontdimen4\font\relax}
\providecommand{\BIBforeignlanguage}[2]{{%
\expandafter\ifx\csname l@#1\endcsname\relax
\typeout{** WARNING: IEEEtranN.bst: No hyphenation pattern has been}%
\typeout{** loaded for the language `#1'. Using the pattern for}%
\typeout{** the default language instead.}%
\else
\language=\csname l@#1\endcsname
\fi
#2}}
\providecommand{\BIBdecl}{\relax}
\BIBdecl

\bibitem[Dakin(1956)]{DakinChapter}
T.~W. Dakin, ``Composite insulating materials and systems,'' in
  \emph{Electrical Properties of Solid Insulating Materials: Molecular
  Structure and Electrical Behavior}, ser. Engineering Dielectrics.\hskip 1em
  plus 0.5em minus 0.4em\relax Baltimore, MD: American Society for Testing and
  Materials, 1956, vol. IIA, pp. 663--695.

\bibitem[Milton(2002)]{MiltonBook}
G.~W. Milton, \emph{The Theory of Composites}, ser. Cambridge monographs on
  applied and computational mathematics.\hskip 1em plus 0.5em minus 0.4em\relax
  Cambridge UK: Cambridge University Press, 2002, vol.~6.

\bibitem[Sihvola(1999)]{SihvolaBook}
A.~Sihvola, \emph{Electromagnetic Mixing Formulas and Applications}, ser. IEE
  Electromagnetic Waves Series.\hskip 1em plus 0.5em minus 0.4em\relax London:
  The Institute of Electrical Engineers, 1999, vol.~47.

\bibitem[Sahimi(2003{\natexlab{a}})]{SahimiBookI}
M.~Sahimi, \emph{Heterogeneous Materials I: Linear Transport and Optical
  Properties}.\hskip 1em plus 0.5em minus 0.4em\relax Berlin: Springer-Verlag,
  2003, vol.~22.

\bibitem[Sahimi(2003{\natexlab{b}})]{SahimiBookII}
------, \emph{Heterogeneous Materials II: Nonlinear and Breakdown Properties
  and Atomistic Modeling}.\hskip 1em plus 0.5em minus 0.4em\relax
  Springer-Verlag, 2003, vol.~23.

\bibitem[Torquato(2001)]{TorquatoBook}
S.~Torquato, \emph{Random Heterogeneous Materials: Microstructure and
  Macroscopic Properties}.\hskip 1em plus 0.5em minus 0.4em\relax Berlin:
  Springer-Verlag, 2001, vol.~16.

\bibitem[Bergman and Stroud(1992)]{Bergman1992}
D.~J. Bergman and D.~Stroud, ``The physical properties of macroscopically
  inhomogeneous media,'' \emph{Solid State Physics}, vol.~46, p. 147, 1992.

\bibitem[Brosseau and Beroual(2003)]{BBReview}
C.~Brosseau and A.~Beroual, ``Computational electromagnetics and the rational
  design of new dielectric heterostructures,'' \emph{Progresses in Materials
  Science}, vol.~48, pp. 373--456, 2003.

\bibitem[Steeman and van Turnhout(2003)]{SteemanTurnhout}
P.~A.~M. Steeman and J.~van Turnhout, ``Dielectric properties of heterogeneous
  media,'' in \emph{Broadband Dielectric Spectroscopy}, F.~Kremer and
  A.~Sch{\"o}nhals, Eds.\hskip 1em plus 0.5em minus 0.4em\relax Berlin:
  Springer-Verlag, 2003, pp. 495--522.

\bibitem[Boyd et~al.(1996)Boyd, Gehr, Fischer, and Sipe]{Boyd1996}
R.~W. Boyd, R.~J. Gehr, G.~L. Fischer, and J.~E. Sipe, ``Nonlinear optical
  properties of nanocomposite materials,'' \emph{Pure Appl. Opt.}, vol.~5, pp.
  505--512, 1996.

\bibitem[Tuncer et~al.(2002{\natexlab{a}})Tuncer, Serdyuk, and
  Gubanski]{Tuncer2002a}
E.~Tuncer, Y.~V. Serdyuk, and S.~M. Gubanski, ``Dielectric mixtures: electrical
  properties and modeling,'' \emph{IEEE Trans. Dielect. Elect. Insul.}, vol.~9,
  no.~5, pp. 809--828, 2002.

\bibitem[Tuncer(2005{\natexlab{a}})]{TuncerPhilMagLett}
E.~Tuncer, ``A formula for dielectric mixtures,'' \emph{Phil. Mag. Lett.},
  vol.~85, no.~6, pp. 269--275, 2005, ({\em Preprint} cond-mat/0503710).

\bibitem[Niklasson and Granqvist(1984)]{niklasson:3382}
G.~A. Niklasson and C.~G. Granqvist, ``Optical properties and solar selectivity
  of coevaporated co-al[sub 2]o[sub 3] composite films,'' \emph{Journal of
  Applied Physics}, vol.~55, no.~9, pp. 3382--3410, 1984.

\bibitem[Clerc et~al.(1990)Clerc, Giraud, Laugier, and Luck]{Clerc1990}
J.~P. Clerc, G.~Giraud, J.~M. Laugier, and J.~M. Luck, ``The electrical
  conductivity of binary disordered systems, percolation clusters, fractals and
  related models,'' \emph{Advances in Physics}, vol.~39, pp. 191--308, 1990.

\bibitem[Nelson(1983)]{NelsonChapter}
K.~Nelson, \emph{Electrical Properties of Solid Insulating Materials: Molecular
  Structure and Electrical Behavior}, ser. Engineering Dielectrics.\hskip 1em
  plus 0.5em minus 0.4em\relax Baltimore, MD: American Society for Testing and
  Materials, 1983, vol. IIA, ch. Breakdown Strength of Solids, pp. 445--520.

\bibitem[Dakin(2006)]{Dakin2006}
T.~Dakin, ``Conduction and polarization mechanisms and trends in dielectric,''
  \emph{Electrical Insulation Magazine, IEEE}, vol.~22, no.~5, pp. 11--28,
  2006.

\bibitem[Berberich and Dakin(1956)]{Dakin1956}
L.~J. Berberich and T.~Dakin, ``Guiding principles in the thermal evaluation of
  electrical insulation,'' \emph{Power Apparatus and Systems, Part III.
  Transactions of the American Institute of Electrical Engineers}, vol.~75,
  no.~3, pp. 752--761, 1956.

\bibitem[Dakin(1978)]{Dakin1978}
T.~W. Dakin, ``High voltage insulation applications,'' \emph{Electrical
  Insulation, IEEE Transactions on}, vol. EI-13, no.~4, pp. 318--326, 1978.

\bibitem[Tuncer(2001)]{TuncerPhD}
E.~Tuncer, ``Dielectric relaxation in dielectric mixtures,'' Ph.D.
  dissertation, Chalmers University of Technology, Gothenburg, Sweden, 2001.

\bibitem[Tuncer et~al.(2001{\natexlab{a}})Tuncer, Guba{\'n}ski, and
  Nettelblad]{Tuncer2001a}
E.~Tuncer, S.~M. Guba{\'n}ski, and B.~Nettelblad, ``Dielectric relaxation in
  dielectric mixtures: {A}pplication of the finite element method and its
  comparison with mixture formulas,'' \emph{J. Appl. Phys.}, vol.~89, no.~12,
  pp. 8092--8100, 2001.

\bibitem[Tuncer et~al.(2011)Tuncer, Belattar, Achour, and
  Brosseau]{TuncerINTECH2011}
E.~Tuncer, J.~Belattar, M.~E. Achour, and C.~Brosseau, ``Broadband spectral
  analysis of non-debye dielectric relaxation in percolating
  heterostructures,'' in \emph{Advances in Composite Materials for Medicine and
  Nanotechnology}, B.~Attaf, Ed.\hskip 1em plus 0.5em minus 0.4em\relax InTech,
  2011, pp. 1--12, doi: 10.5772/14680.

\bibitem[Tuncer et~al.(2002{\natexlab{b}})Tuncer, Nettelblad, and
  Guba{\'n}ski]{Tuncer2002b}
E.~Tuncer, B.~Nettelblad, and S.~M. Guba{\'n}ski, ``Non-debye dielectric
  relaxation in binary dielectric mixtures (50-50): {R}andomness and regularity
  in mixture topology,'' \emph{J. Appl. Phys.}, vol.~92, no.~8, pp. 4612--4624,
  2002.

\bibitem[Tuncer(2010)]{ma3010585}
\BIBentryALTinterwordspacing
E.~Tuncer, ``Geometrical description in binary composites and spectral density
  representation,'' \emph{Materials}, vol.~3, no.~1, pp. 585--613, 2010.
  [Online]. Available: \url{http://www.mdpi.com/1996-1944/3/1/585}
\BIBentrySTDinterwordspacing

\bibitem[Tuncer(2005{\natexlab{b}})]{TuncerJPD2005}
------, ``Structure/property relationship in dielectric mixtures: application
  of the spectral density theory,'' \emph{J. Phys. D:Appl. Phys.}, vol.~38, pp.
  223--234, 2005.

\bibitem[Tuncer(2005{\natexlab{c}})]{TuncerSpectralPRB}
------, ``Extracting spectral density function of a binary composite without
  a-priori assumption,'' \emph{Phys. Rev. B}, vol.~71, p. 012101, 2005, ({\it
  Preprint} cond-mat/0403243).

\bibitem[Tuncer(2005{\natexlab{d}})]{Tuncer2005JPCMLET}
------, ``Analogy between dielectric relation and dielectric mixtures:
  Application of the spectral density representation,'' \emph{J. Phys.:
  Condens. Matter}, vol.~17, no.~12, pp. L125--L128, 2005, ({\em Preprint}
  cond-mat/0502580).

\bibitem[Eyre and Milton(1999)]{EyreMilton}
D.~J. Eyre and G.~W. Milton, ``A fast numerical scheme for computing the
  response of composites using grid refinement,'' \emph{The European Physical
  Journal Applied Physics}, vol.~6, pp. 41--47, 1999.

\bibitem[Milton et~al.(1981)Milton, McPhedran, and McKenzie]{mil81}
G.~Milton, R.~McPhedran, and D.~McKenzie, ``Transport properties of arrays of
  intersecting cylinders,'' \emph{Applied Physics}, vol.~25, pp. 23--40, 1981.

\bibitem[Brosseau and Beroual(2001)]{Brosseau1}
C.~Brosseau and A.~Beroual, ``Effective permittivity of composites with
  stratified particles,'' \emph{Journal of Physics D: Applied Physics},
  vol.~34, pp. 704--710, 2001.

\bibitem[Brosseau and Beroual(1999)]{Brosseau2}
------, ``Dielectric properties of periodic heterostructures: {A} computational
  electrostatics approach,'' \emph{The European Physical Journal Applied
  Physics}, vol.~6, pp. 23--31, 1999.

\bibitem[Sareni et~al.(1997{\natexlab{a}})Sareni, Kr{\"a}henb{\"u}hl, Beroual,
  Nicolas, and Brosseau]{Sareni1}
B.~Sareni, L.~Kr{\"a}henb{\"u}hl, A.~Beroual, A.~Nicolas, and C.~Brosseau, ``An
  {I}ntio simulation approach for calculating the effective dielectric constant
  of composite materials,'' \emph{Journal of Electrostatics}, vol. 40 {\&} 41,
  pp. 489--494, 1997.

\bibitem[Sareni et~al.(1997{\natexlab{b}})Sareni, Kr{\"a}henb{\"u}hl, Beroual,
  and Brosseau]{sar97}
B.~Sareni, L.~Kr{\"a}henb{\"u}hl, A.~Beroual, and C.~Brosseau, ``Effective
  dielectric constant of random composite materials,'' \emph{J. Appl. Phys.},
  vol.~81, no.~5, pp. 2375--2383, 1997.

\bibitem[Sareni et~al.(1997{\natexlab{c}})Sareni, Kr{\"a}henb{\"u}hl, Beroual,
  Nicolas, and Brosseau]{sar97mag}
B.~Sareni, L.~Kr{\"a}henb{\"u}hl, A.~Beroual, A.~Nicolas, and C.~Brosseau, ``A
  boundary integral equation method for calculation of the effective
  permittivity of periodic composites,'' \emph{IEEE Transactions on Magnetics},
  vol.~33, no.~2, pp. 7240--7246, 1997.

\bibitem[Maxwell(1891)]{Maxwellbook}
J.~C. Maxwell, \emph{A Treatise on Electricity and Magnetism- Volume 1},
  3rd~ed.\hskip 1em plus 0.5em minus 0.4em\relax Oxford: Clarendon Press, 1891,
  reprint by Dover.

\bibitem[Scaife(1998)]{Scaife}
B.~K.~P. Scaife, \emph{Principles of Dielectrics}.\hskip 1em plus 0.5em minus
  0.4em\relax Oxford Science Publications, 1998.

\bibitem[Wagner(1913)]{Wagner1913}
K.~W. Wagner, ``Zur theorie der unvollkommenen dielektrika,'' \emph{Annalen der
  Physik}, vol.~40, no.~5, pp. 817--855, 1913.

\bibitem[Fricke(1924)]{Fricke1924}
H.~Fricke, ``A mathematical treatment of the electric conductivity and capacity
  of disperse systems,'' \emph{Physical Review}, vol.~24, pp. 575--587, 1924.

\bibitem[Sillars(1937)]{Sillars1937}
R.~Sillars, ``The properties of a dielectric containing semiconducting
  particles of various shapes,'' \emph{Journal of Institution of Electrical
  Engineers}, vol.~80, pp. 378--394, 1937.

\bibitem[Fuchs(1975)]{Fuchs1975}
R.~Fuchs, ``Theory of the optical properties of ionic crystal cubes,''
  \emph{Phys. Rev.}, vol. B11, p. 1732, 1975.

\bibitem[Bergman(1978)]{Bergman1978}
D.~J. Bergman, ``The dielectric constant of a composite material---a problem in
  classical physics,'' \emph{Physics Reports}, vol.~43, no.~9, pp. 377--407,
  1978.

\bibitem[Bergman(1979{\natexlab{a}})]{Bergman1979}
D.~Bergman, ``The dielectric constant of a simple cubic array of identical
  spheres,'' \emph{Journal of Physics C: Solid State Physics}, vol.~12, pp.
  4947--4960, 1979.

\bibitem[Bergman(1980)]{Bergman3}
D.~J. Bergman, ``Exactly solvable microscopic geometries and rigorous bounds
  for the complex dielectric constant of a two-component composite material,''
  \emph{Phys. Rev. Lett.}, vol.~44, no.~19, pp. 1285--1287, 1980.

\bibitem[Bergman(1982)]{bergman1982}
------, ``Rigorous bounds for complex dielectric constant of a two-component
  composite,'' \emph{Annals of Physics}, vol. 138, pp. 78--114, 1982.

\bibitem[Bergman and Dunn(1992)]{Bergman1}
D.~J. Bergman and K.-J. Dunn, ``Bulk effective dielectric constant of a
  composite with a periodic microgeometry,'' \emph{Phys. Rev. B}, vol.~45,
  no.~23, pp. 13\,262--13\,271, 1992.

\bibitem[Milton(1981{\natexlab{a}})]{Miltona}
G.~W. Milton, ``Bounds on the electromagnetic, elastic, and other properties of
  two-component composites,'' \emph{Phys. Rev. Lett.}, vol.~46, no.~8, pp.
  542--545, 1981.

\bibitem[Milton(1981{\natexlab{b}})]{Milton1981b}
------, ``Bounds on the transport and optical properties of a two-component
  composite material,'' \emph{J. Appl. Phys.}, vol.~52, no.~8, pp. 5294--5304,
  1981.

\bibitem[Milton(1981{\natexlab{c}})]{Milton1981}
------, ``Bounds on the complex permittivity of a two-component composite
  material,'' \emph{J. Appl. Phys.}, vol.~52, pp. 5286--5293, 1981.

\bibitem[Golden and Papanicolaou(1983)]{Golden1}
K.~Golden and G.~Papanicolaou, ``Bounds on effective parameters of
  heterogeneous media by analytic continuation,'' \emph{Commun. Math. Phys.},
  vol.~90, pp. 473--491, 1983.

\bibitem[Golden and Papanicolaou(1985)]{Golden2}
------, ``Bounds for effective parameters of multicomponent media by analytical
  continuation,'' \emph{J. Stat. Phys.}, vol.~40, no. 4/5, pp. 655--667, 1985.

\bibitem[Cherkaev and Zhang(2003)]{Cherkaev2003}
E.~Cherkaev and D.~Zhang, ``Coupling of the efective properties of a random
  mixture through the reconsructed spectral representation,'' \emph{Physica B},
  vol. 338, pp. 16--23, 2003.

\bibitem[Day et~al.(2000{\natexlab{a}})Day, Grant, Sievers, and
  Thorpe]{DayPRL2000}
A.~R. Day, A.~R. Grant, A.~J. Sievers, and M.~F. Thorpe, ``Spectral function of
  composites from reflectivity measurements,'' \emph{Phys. Rev. Lett.},
  vol.~84, no.~9, pp. 1978--1981, 2000.

\bibitem[Day et~al.(2000{\natexlab{b}})Day, Thorpe, Grant, and
  Sievers]{DayPhysB1}
A.~R. Day, M.~F. Thorpe, A.~G. Grant, and A.~J. Sievers, ``The spectral
  function of a composite from reflectance data,'' \emph{Physica B}, vol. 279,
  pp. 17--20, 2000.

\bibitem[Day et~al.(2000{\natexlab{c}})Day, McGrun, Bergman, and
  Thorpe]{DayPhysB2}
A.~R. Day, A.~R. McGrun, D.~J. Bergman, and M.~F. Thorpe, ``The spectral
  function of the electrical properties of layered materials,'' \emph{Physica
  B}, vol. 338, pp. 24--30, 2000.

\bibitem[Day and Thorpe(1999)]{Day}
A.~R. Day and M.~F. Thorpe, ``The spectral fuction of composites,'' \emph{J.
  Phys.: Condens. Matter}, vol.~11, pp. 2551--2568, 1999.

\bibitem[Barabash and Stroud(1999)]{Stroud1999}
S.~Barabash and D.~Stroud, ``Spectral representation for the effective
  macroscopic response of a polycrystal: application to third-order non-linear
  susceptibility,'' \emph{J. Phys.: Condens. Matter}, vol.~11, pp.
  10\,323--10\,334, 1999.

\bibitem[Fuchs and Claro(1989)]{Fuchs_a}
R.~Fuchs and F.~Claro, ``Spectral representation for the polarizability of a
  collection of dielectric spheres,'' \emph{Phys. Rev. B}, vol.~39, no.~6, pp.
  3875--3878, 1989.

\bibitem[Ghosh and Fuchs(1988)]{GhoshFuchs}
K.~Ghosh and R.~Fuchs, ``Spectral theory of two-component porous media,''
  \emph{Phys. Rev. B}, vol.~38, no.~8, pp. 5222--5236, 1988.

\bibitem[Pecharrom\'an and Gordillo-V\'azquez(2006)]{Pech_PhysRevB.74.035120}
C.~Pecharrom\'an and F.~J. Gordillo-V\'azquez, ``Expansion of the spectral
  representation function of a composite material in a basis of legendre
  polynomials: Experimental determination and analytic approximations,''
  \emph{Phys. Rev. B}, vol.~74, p. 035120, 2006.

\bibitem[Wiener(1912)]{wiener}
O.~Wiener, ``Die {T}heorie des {M}ischk{\"o}rpers f{\"u}r das {F}eld der
  staton{\"a}ren {S}tr{\"o}mung {I}. {D}ie {M}ittelwerts{\"a}tze f{\"u}r
  {K}raft, {P}olarisation und {E}nergie,'' \emph{Der Abhandlungen der
  Mathematisch-Physischen Klasse der K{\"o}nigl. Sachsischen Gesellschaft der
  Wissenschaften}, vol.~32, pp. 509--604, 1912.

\bibitem[Bruggeman(1935)]{Bruggeman1935}
D.~A.~G. Bruggeman, ``Berechnung verschiedener physikalischer {K}onstanten von
  heterogenen {S}ubstanzen,'' \emph{Annalen der Physik (Leipzig)}, vol.~24, pp.
  636--679, 1935.

\bibitem[Widjajakusuma et~al.(2003)Widjajakusuma, Biswal, and
  Hilfer]{Widjajakusuma2003319}
J.~Widjajakusuma, B.~Biswal, and R.~Hilfer, ``Quantitative comparison of mean
  field mixing laws for conductivity and dielectric constants of porous
  media,'' \emph{Physica A: Statistical Mechanics and its Applications}, vol.
  318, no. 3–4, pp. 319 -- 333, 2003.

\bibitem[Boudida et~al.(2000)Boudida, Beroual, and Brosseau]{Boudida2}
A.~Boudida, A.~Beroual, and C.~Brosseau, ``How do shape anisotropy and spatial
  orientation of the constiruents affect the permittivity of dielectric
  heterostructures,'' \emph{J. Appl. Phys.}, vol.~88, no.~12, pp. 7278--7288,
  2000.

\bibitem[Boudida et~al.(1998)Boudida, Beroual, and Brosseau]{Boudida}
------, ``Permittivity of lossy composite materials,'' \emph{J. Appl. Phys.},
  vol.~83, no.~1, pp. 425--431, 1998.

\bibitem[Mejdoubi and Brosseau(2006)]{BrosseauPhysRevE.74.031405}
A.~Mejdoubi and C.~Brosseau, ``Finite-element simulation of the depolarization
  factor of arbitrarily shaped inclusions,'' \emph{Phys. Rev. E}, vol.~74, p.
  031405, 2006.

\bibitem[Myroshnychenko and Brosseau(2005)]{BrosseauPhysRevE71}
V.~Myroshnychenko and C.~Brosseau, ``Finite-element method for calculation of
  the effective permittivity of random inhomogeneous media,'' \emph{Phys. Rev.
  E}, vol.~71, p. 016701, 2005.

\bibitem[Serdyuk et~al.(2004)Serdyuk, Podoltsev, and Gubanski]{Serdyuk1306717}
Y.~Serdyuk, A.~Podoltsev, and S.~Gubanski, ``Numerical simulations and
  experimental study of frequency-dependent dielectric properties of composite
  material with stochastic structure,'' \emph{Dielectrics and Electrical
  Insulation, IEEE Transactions on}, vol.~11, no.~3, pp. 379--392, 2004.

\bibitem[Kühn and Kliem(2010)]{Kliem5595551}
M.~Kühn and H.~Kliem, ``A numerical method for the calculation of dielectric
  nanocomposites,'' \emph{Dielectrics and Electrical Insulation, IEEE
  Transactions on}, vol.~17, no.~5, pp. 1499--1508, 2010.

\bibitem[Beroual and Brosseau(2001)]{Beroual971447}
A.~Beroual and C.~Brosseau, ``Comparison of dielectric properties determined
  from a computational approach and experiment for anisotropic and periodic
  heterostructures,'' \emph{Dielectrics and Electrical Insulation, IEEE
  Transactions on}, vol.~8, no.~6, pp. 921--929, 2001.

\bibitem[Blanchard et~al.(2007)Blanchard, Porti, Morente, Salinas, and
  Navarro]{blanchard:064101}
C.~Blanchard, J.~A. Porti, J.~A. Morente, A.~Salinas, and E.~A. Navarro,
  ``Determination of the effective permittivity of dielectric mixtures with the
  transmission line matrix method,'' \emph{Journal of Applied Physics}, vol.
  102, no.~6, 2007.

\bibitem[Murugaraj et~al.(2005)Murugaraj, Mainwaring, and
  Mora-Huertas]{murugaraj:054304}
P.~Murugaraj, D.~Mainwaring, and N.~Mora-Huertas, ``Dielectric enhancement in
  polymer-nanoparticle composites through interphase polarizability,''
  \emph{Journal of Applied Physics}, vol.~98, no.~5, 2005.

\bibitem[K{\"a}rk{\"a}inen et~al.(2001)K{\"a}rk{\"a}inen, Sihvola, and
  Nikoskinen]{Karkkainen}
K.~K{\"a}rk{\"a}inen, A.~Sihvola, and K.~Nikoskinen, ``Analysis of a
  three-dimensional dielectric mixture with the finite difference method,''
  \emph{IEEE Transactions on Geoscience and Remote Sensing}, vol.~39, no.~5,
  2001.

\bibitem[Avelin et~al.(2001)Avelin, Sharma, H{\"a}nninen, and
  Sihvola]{SihvolaPol}
J.~Avelin, R.~Sharma, I.~H{\"a}nninen, and A.~Sihvola, ``Polarizability
  analysis of cubical and square-shaped dielectric scatterers,'' \emph{IEEE
  Transactions on Antennas and Propagation}, vol.~49, no.~3, pp. 451--457,
  2001.

\bibitem[K{\"a}rkk{\"a}inen et~al.(2000)K{\"a}rkk{\"a}inen, Sihvola, and
  Nikoskinen]{SihvolaEff}
K.~K. K{\"a}rkk{\"a}inen, A.~Sihvola, and K.~I. Nikoskinen, ``Effective
  permittivity of mixtures: {N}umerical validation by the fdtd method,''
  \emph{IEEE Transactions on Geoscience and Remote Sensing}, vol.~38, no.~3,
  pp. 1303--1308, 2000.

\bibitem[Pekonen et~al.(1999)Pekonen, K{\"a}rk{\"a}inen, A, and
  Nikoskinen]{Pekonen1999}
O.~Pekonen, K.~K{\"a}rk{\"a}inen, S.~A, and K.~Nikoskinen, ``Numerical testing
  of dielectric mixing rules by {FDTD} method,'' \emph{Journal of
  Electromagnetic Waves and Applications}, vol.~13, pp. 67--87, 1999.

\bibitem[Sihvola and Lindell(1990)]{Sihvola-ellips}
A.~Sihvola and I.~V. Lindell, ``Polarizability and effective permittivity of
  layered and discontinuously inhomogeneous dielectric ellipsoids,''
  \emph{Journal of Electromagnetic Waves and Applications}, vol.~4, no.~1, pp.
  1--26, 1990.

\bibitem[Tuncer(2003)]{TuncerAcc1}
E.~Tuncer, ``How round is round? on accuracy in complex dielectric permittivity
  calculations: A finite-size scaling approach,'' \emph{{T}urkish Journal of
  Physics}, vol.~27, pp. 121--131, 2003, {\tt arXiv:cond-mat/0107384}.

\bibitem[K{\"a}rk{\"a}inen(2002)]{KarkkainenPhD}
K.~K{\"a}rk{\"a}inen, ``On the finite-difference modelling of electromagnetic
  problems in structured lattices,'' Electromagnetics Laboratory Report Series,
  Rep. 403, Helsinki University of Technology, Helsinki Finland, 2002.

\bibitem[Tuncer et~al.(2001{\natexlab{b}})Tuncer, Guba{\'n}ski, and
  Nettelblad]{Tuncer2002elec}
E.~Tuncer, S.~M. Guba{\'n}ski, and B.~Nettelblad, ``Electrical properties of
  $4\times4$ binary dielectric mixtures,'' \emph{Journal of Electrostatics},
  vol.~56, no.~4, pp. 449--463, 2001.

\bibitem[Clauzon et~al.(1999)Clauzon, Kr{\"a}henb{\"u}hl, and Nicolas]{Clauzon}
P.~Clauzon, L.~Kr{\"a}henb{\"u}hl, and A.~Nicolas, ``Effective permittivity of
  3{D} lossy dielectric composite materials,'' \emph{IEEE Transactions on
  Magnetics}, vol.~35, no.~3, pp. 1223--1226, 1999.

\bibitem[Merrill et~al.(1999)Merrill, Diaz, LoRe, Squires, and
  Alexopoulos]{Merrill}
W.~Merrill, R.~E. Diaz, M.~M. LoRe, M.~C. Squires, and N.~G. Alexopoulos,
  ``Effective medium theories for artificial materials composed of multiple
  sizes of spherical inclusions in a host medium,'' \emph{IEEE Transactions on
  Antennas and Propagation}, vol.~47, no.~1, pp. 142--148, 1999.

\bibitem[Lui and Xu(1997)]{CLiu}
C.~Lui and H.~Xu, ``Computation of the effective dielectric constant of
  two-component, three-dimensional mmixtures using a simple pole expansion
  method,'' \emph{J. Appl. Phys.}, vol.~82, no.~1, pp. 345--350, 1997.

\bibitem[Liu and Wu(1997)]{liu97}
C.~Liu and H.~Wu, ``Computation of the effective dielectric constant of
  two-component, three dimensional mixtures using simple pole expansion
  method,'' \emph{J. Appl. Phys.}, vol.~82, no.~1, pp. 345--350, 1997.

\bibitem[Almond and Vainas(1999)]{Vainas}
D.~P. Almond and B.~Vainas, ``The dielectric properties of random {R}-{C}
  networks as an explanation of the 'universal' power law dielectric response
  of solids,'' \emph{J. Phys.: Condens. Matter}, vol.~11, pp. 9081--9093, 1999.

\bibitem[Hamou et~al.(2009)Hamou, Macdonald, and Tuncer]{TuncerFaycalRoss2009}
R.~F. Hamou, J.~R. Macdonald, and E.~Tuncer, ``Dispersive dielectric and
  conductive effects in 2d resistor\&ndash;capacitor networks,'' \emph{Journal
  of Physics: Condensed Matter}, vol.~21, no.~2, p. 025904 (13pp), 2009.

\bibitem[Gershon et~al.(2001)Gershon, Calame, and Birnboim]{Calame2001JAP}
D.~Gershon, J.~P. Calame, and A.~Birnboim, ``Complex permittivity measurements
  and mixing laws of alumina composites,'' \emph{J. Appl. Phys.}, vol.~89,
  no.~12, pp. 8110--8116, 2001.

\bibitem[Calame(2003)]{Calame2003}
J.~P. Calame, ``Evolution of davidson-cole relaxation behavior in random
  conductor-insulator composites,'' \emph{J. Appl. Phys.}, vol.~94, no.~9, pp.
  5945--5957, 2003.

\bibitem[An et~al.(2008)An, Boggs, and Calame]{Boggs4591430}
L.~An, S.~Boggs, and J.~Calame, ``Energy storage in polymer films with high
  dielectric constant fillers,'' \emph{Electrical Insulation Magazine, IEEE},
  vol.~24, no.~3, pp. 5--10, 2008.

\bibitem[Fornes and Paul(2003)]{Fornes20034993}
T.~Fornes and D.~Paul, ``Modeling properties of nylon 6/clay nanocomposites
  using composite theories,'' \emph{Polymer}, vol.~44, no.~17, pp. 4993 --
  5013, 2003.

\bibitem[Priou(1992)]{Pier6}
A.~Priou, Ed., \emph{Progress in Electromagnetics Research}, ser. Dielectric
  Properties of Heterogeneous Materials.\hskip 1em plus 0.5em minus 0.4em\relax
  New York: Elsevier, 1992.

\bibitem[Tuncer(2012)]{TuncerSpectralAPAMSP}
E.~Tuncer, ``Spectral density representation of dielectric mixtures,''
  \emph{Applied Physics A}, vol. 107, pp. 575--582, 2012.

\bibitem[Fuchs and Liu(1976)]{FuchsLiu}
R.~Fuchs and S.~H. Liu, ``Sum rule for polarizability of small particles,''
  \emph{Phys. Rev. B}, vol.~14, no.~12, pp. 5521--5522, 1976.

\bibitem[Bergman(1979{\natexlab{b}})]{Bergman4}
D.~J. Bergman, ``Dielectric constant of a two-component granular composite: {A}
  practical scheme for calculating the pole spectrum,'' \emph{Phys. Rev. B},
  vol.~19, no.~4, pp. 2359--2368, 1979.

\bibitem[Stauffer and Aharony(1991)]{Percolation}
D.~Stauffer and A.~Aharony, \emph{Introduction to Percolation Theory}, revised
  2nd~ed.\hskip 1em plus 0.5em minus 0.4em\relax London: Taylor {\&} Francis,
  1991.

\bibitem[Ghosh and Fuchs(1991)]{Ghosh}
K.~Ghosh and R.~Fuchs, ``Critical behavior in the dielectric properties of
  random self-similar composites,'' \emph{Phys. Rev. B}, vol.~44, no.~14, pp.
  7330--7343, 1991.

\bibitem[Tuncer and Niklasson(2008)]{Tuncer2008OptComm}
E.~Tuncer and G.~A. Niklasson, ``Optical properties of non-dilute
  metal-insulator composites,'' \emph{Optics Communications}, vol. 281, no.~17,
  pp. 4374--4379, 2008.

\bibitem[Tuncer and Drummy(2010)]{TuncerDrummy}
E.~Tuncer and L.~Drummy, ``Novel numerical method for acquiring a geometrical
  description of nanodielectrics,'' in \emph{Power Modulator and High Voltage
  Conference (IPMHVC), 2010 IEEE International}, 2010, pp. 42--44.

\bibitem[Milton(1980)]{milton:300}
G.~W. Milton, ``Bounds on the complex dielectric constant of a composite
  material,'' \emph{Applied Physics Letters}, vol.~37, no.~3, pp. 300--302,
  1980.

\bibitem[Hashin and Shtrikman(1962)]{hashin}
Z.~Hashin and S.~Shtrikman, ``A variational approach to the theory of effective
  magnetic pereability of multiphase materials,'' \emph{J. Appl. Phys.},
  vol.~33, pp. 3125--3131, 1962.

\end{thebibliography}
\bibliographystyle{IEEEtranN}
\end{document}